*Word count: 4228*

# Insights into the long-standing controversy over sound velocities in lizardite

Chenxing Luo[1] (ORCID: 0000-0003-4116-6851), Timothy Liao[2,3] (ORCID: 0000-0001-6175-7734), Hongjin Wang[1] (ORCID: 0009-0008-9766-1177), and Renata M. Wentzcovitch[1,2,3,*] (ORCID: 0000-0001-5663-9426)

*1 Department of Applied Physics and Applied Mathematics, Columbia University, New York, New York 10027, USA*

*2 Department of Earth and Environmental Sciences, Columbia University, New York, New York 10027, USA*

*3 Lamont–Doherty Earth Observatory, Columbia University, Palisades, New York 10964, USA*

* rmw2150@columbia.edu

***Abstract.*** Serpentine minerals are abundant H-bearing phases in the upper mantle and play a central role in water transport in subduction zones. Lizardite is the low-temperature polymorph expected in cold slabs and shallow serpentinized mantle. Using an ab initio thermoelastic framework combining density-functional theory with the r²SCAN meta-GGA and the quasiharmonic approximation, we calculate its compression curve, elastic tensor, and acoustic velocities at 300 K. r²SCAN reproduces the measured compression curve and equilibrium volume to within 0.2%. Weak interlayer hydrogen bonding gives rise to pronounced axial softening, with $C_{33} \approx 0.21 C_{11}$ and $C_{44} \approx 0.13 C_{66}$ at 0 GPa. Solving the Christoffel equation over all propagation directions, we construct the velocity density of states (VelDOS), which reveals a high density of slow acoustic modes and other directional features obscured by Voigt–Reuss–Hill (VRH) aggregate averages. Reported experimental velocities preferentially sample the slower portion of this distribution rather than clustering near the VRH values, a correspondence consistent with subtle crystallographic texture or preferential orientation not resolved experimentally. Thus, the appropriate seismic-velocity endmember for lizardite depends on its orientational state, and reliance on VRH velocities alone may lead to overestimation of the degree of serpentinization when seismic observations preferentially sample slow crystallographic directions. Direction-resolved single-crystal measurements and aggregate measurements with independently quantified texture would provide direct tests of this interpretation.

## INTRODUCTION

Serpentines (Evans 2004) are among the most abundant H-bearing minerals in the upper mantle. Their layered structures and high water contents make them central to hydration, dehydration, and water transport in subduction zones. Among the serpentine polymorphs, lizardite is characteristic of low-temperature hydration and cold subduction environments.

Lizardite has a chemical formula of $Mg_3Si_2O_5(OH)_4$ and trigonal symmetry, with space group P31m (No. 157) (Mellini and Zanazzi 1987; Tsuchiya 2013). Its structure consists of Mg–O octahedral sheets bonded to Si–O tetrahedral sheets, forming flat layers stacked along the crystallographic *c* direction (Fig. 1). Neighboring layers are connected primarily by H-bonds, so compression within the basal *a*-*b* plane is governed mainly by stronger Mg–O and Si–O bonding, whereas compression along *c* samples the softer interlayer H-bond network. This contrast between stiff intralayer bonding and softer interlayer bonding is the structural origin of lizardite's strong elastic anisotropy.

Serpentines are challenging for *ab initio* simulations because their physical properties are directly determined by stoichiometric H-bonds, whose compressive behaviors are not well described by standard density-functional theory (DFT) functionals such as the local-density approximation (LDA) (Perdew and Zunger 1981) and the Perdew-Burke-Ernzerhof (PBE) generalized-gradient approximation (GGA) (Perdew et al. 1996). In addition, semilocal functionals can overestimate or underestimate pressure. The strongly constrained and appropriately normed ($r^2$SCAN) meta-GGA functional (Sun et al. 2016) more accurately describes the structures and energetics of systems with mixed covalent, ionic, van der Waals, and H-bonding interactions. We therefore use $r^2$SCAN to establish an accurate 300 K structural baseline before evaluating lizardite's elastic anisotropy and acoustic velocities.

Measurements on lizardite/chrysotile-bearing low-temperature serpentinites have long established these rocks as low-velocity materials, with estimates of $V_P \sim 4.8$ km/s and $V_S \sim 2.3$ km/s at several hundred MPa to 1 GPa widely cited in the geophysical literature (Christensen 1966; Christensen 2004; Ji et al. 2013). Watanabe et al. (2007) measured wave velocities at room temperature and confining pressures up to 200 MPa in OHM-B, a low-temperature serpentinite in which lizardite and chrysotile together account for 95 vol% of the sample. No foliation was observed. At 200 MPa, the arithmetic mean of the three $V_P$ measurements was 5.03 km/s, and the corresponding mean $V_S$ was 2.62 km/s. The three $V_P$ values, 5.05, 5.02, and 5.01 km/s, differed only slightly, consistent with weak velocity anisotropy at the scale of the rock sample. By contrast, previous first-principles calculations of lizardite typically give higher isotropic-average velocities, with $V_P \sim$ 6–7 km/s and $V_S \sim$ 3.5–4.0 km/s (Deng et al. 2022). This contrast persists across exchange-correlation functionals and thermal treatments: changing from LDA to PBE or including thermal effects in previous calculations does not close the gap with experiment. Sample composition and microstructure may also affect the measured values, including through chrysotile, nanotube-like morphology, porosity, cracks, or other defects (Mookherjee and Stixrude 2009; Reynard et al. 2007). The similarity of values reported for independent samples motivates examining whether the comparison with isotropic aggregate averages omits important directional information.

Here we revisit this problem by analyzing the full directional velocity distribution. Previous computational studies generally compared experimental values with velocities derived from Voigt–Reuss–Hill (VRH) averaged bulk and shear moduli (Hill 1952; Karki et al. 2001). These averages provide useful estimates of an effective isotropic aggregate response, but they reduce lizardite's strong elastic anisotropy to a single $V_P$ and a single $V_S$. The Voigt and Reuss estimates provide bounds on this isotropic response, while the Hill value is their arithmetic mean; none retains the full directional information contained in the elastic tensor. In a layered mineral such as lizardite, the Christoffel equation instead determines the complete set of directional $V_P$, $V_{S1}$, and $V_{S2}$ values. We therefore analyze the full velocity distribution and compare its characteristic velocities with experimental observations. We find that the experimentally reported values are closer to slower portions of the directional distribution than to the conventional VRH averages. Because seismic interpretations treat lizardite velocity as an endmember, this distinction also affects velocity-based estimates of serpentinization and associated water content.

## COMPUTATIONAL METHODS

We compute the thermodynamic and elastic properties of lizardite using *ab initio* methods. DFT calculations were performed with the projector augmented-wave (PAW) method implemented in the Vienna *Ab initio* Simulation Package (VASP) v6.3 (Kresse and Furthmüller 1996) with the r$^2$SCAN meta-GGA functional (Furness et al. 2020). The plane-wave energy cutoff was set to 520 eV. The Brillouin zone was sampled on a shifted 4 × 4 × 4 Monkhorst-Pack *k*-point mesh. Compared to PBE (Perdew et al. 1996) and LDA (Perdew and Zunger 1981), the SCAN meta-GGA was developed to address the "structure or energy" dilemma and to capture intermediate-range van der Waals interactions (Sun et al. 2016), with SCAN/r$^2$SCAN applications describing well the compressive behavior and elastic properties of hydrous minerals, including $\delta$-AlOOH, brucite, serpentine, and hydrous stishovite (Luo et al. 2024; Wang et al. 2024a; Wang et al. 2024b; Zhang and Luo 2026). These interactions matter for lizardite because its strong intralayer Mg–O and Si–O bonds, weaker interlayer hydrogen bonds, and intermediate-range dispersion interactions together determine its equilibrium structure, compression, and elastic response.

Phonon calculations were performed with the PHONOPY package (Togo and Tanaka 2015). The force sets were obtained with a 2 × 2 × 2 supercell and then interpolated to a 9 × 9 × 9 $q$-point mesh for Brillouin-zone sampling. The static elastic tensors were obtained using the stress-strain method (Karki et al. 2001; da Silveira et al. 2013) with a strain amplitude of ± 0.2%. The thermodynamic properties at finite temperature were calculated using the quasiharmonic approximation (QHA) implemented in the QHA package (Qin et al. 2019); the elastic tensors at finite temperature were calculated using the Wu-Wentzcovitch semi-analytical method (Wu and Wentzcovitch 2011) for thermoelastic constants (SAM-Cij) implemented in the CIJ package (Luo et al. 2021).

The thermal elastic coefficients are obtained from the strain derivatives of the Helmholtz free energy (Wu and Wentzcovitch 2011; Luo et al. 2021; Luo et al. 2022),

$$C_{ijkl}^{T} = \frac{1}{V}\left(\frac{\partial^2 F}{\partial e_{ij}\partial e_{kl}}\right) + \frac{1}{2}P\left(2\delta_{ij}\delta_{kl} - \delta_{il}\delta_{jk} - \delta_{ik}\delta_{jl}\right), \quad (1)$$

where $C_{ijkl}^{T}$ is the isothermal elastic coefficient, $V$ is volume, $F$ is the Helmholtz free energy, $e_{ij}$ and $e_{kl}$ are strain components, $P$ is pressure, and $\delta_{ij}$ is the Kronecker delta. Under the QHA, the Helmholtz free energy is (Qin et al. 2019)

$$F(e,V,T) = U^{\text{st}}(e,V) + \frac{1}{3N}\sum_{qm}\frac{1}{2}\hbar\omega_{qm}(e,V) + \frac{k_B T}{3N}\sum_{qm}\ln\left\{1-\exp\left[-\frac{\hbar\omega_{qm}(e_{ij},V)}{k_B T}\right]\right\}. \quad (2)$$

Here $e$ is the strain state, $T$ is temperature, $U^{\text{st}}$ is the static internal energy, $N$ is the number of atoms in the simulation cell, $q$ and $m$ index phonon wave vectors and branches, $\omega_{qm}$ is the phonon frequency, $\hbar$ is the reduced Planck constant, and $k_B$ is the Boltzmann constant. These expressions connect the strain response of the static and vibrational free energy to the finite-temperature elastic tensor.

The crystal structure of lizardite is deposited as a Crystallographic Information File (CIF) accompanying this paper. Three additional figures are also provided as Supplementary Figures S1–S3: the pressure-dependent elastic tensor components (Supplementary Figure S1), the incidence angles of the off-axis velocity maxima $V_P^*$ and $V_S^*$ (Supplementary Figure S2), and a comparison of the LDA and PBE velocity distributions with the r$^2$SCAN results and experiment (Supplementary Figure S3).

## RESULTS AND DISCUSSION

The 300 K compression curves are shown in Fig. 2. The r$^2$SCAN curve closely follows the in situ X-ray diffraction data of Hilairet et al. (2006). The calculated equilibrium volume, $V_0$=180.55 Å$^3$ (Table I), differs from the experimental value of 180.92 Å$^3$ by 0.37 Å$^3$ (0.2%), comparable to the largest residual ( ~0.4 Å$^3$) of the individual diffraction volumes from the fitted experimental equation of state and substantially smaller than the 4.6% LDA underestimate and 2.2% PBE overestimate. The agreement extends to both lattice parameters [Fig. 2(b)], demonstrating that r$^2$SCAN captures compression along both the strongly bonded intralayer *a* direction and the weakly hydrogen-bonded interlayer *c* direction.

Table I also compares the calculated elastic tensor and aggregate velocities with previous calculations and experiments. The tensor reflects the mechanical stratification of lizardite (i.e., stiff Mg–O/Si–O-bonded layers separated by compliant hydrogen-bonded interfaces). At 300 K and 0 GPa, the *c*-axis longitudinal response and shear involving *c* are much softer than their basal-plane counterparts, with $C_{33}\approx0.21C_{11}$ and $C_{44}\approx0.13C_{66}$.

The functional comparison identifies interlayer bonding as the source of the divergent elastic response. The basal-plane stiffness $C_{11}$, governed by strong intralayer bonding, varies comparatively little among functionals. In contrast, the r$^2$SCAN $C_{33}$ is comparable to the softer GGA estimates and is less than half the LDA values; $C_{44}$ shows the same ordering. The larger LDA values correspond to a stiffer description of the hydrogen-bonded interfaces. The r$^2$SCAN tensor predicts pronounced interlayer compliance consistent with their weak bonding.

The resulting elastic anisotropy produces slow *c*-axis acoustic modes because $V_P$[001] and $V_S$[001] depend primarily on $C_{33}$ and $C_{44}$, respectively. VRH averaging condenses these directional modes into one isotropic $V_P$ and $V_S$, obscuring the slow response associated with the compliant interlayer direction. We therefore calculate the full directional distributions of $V_P$, $V_{S1}$, and $V_{S2}$, summarize them with the VelDOS defined below, and compare their characteristic velocities with experiment.

The pressure-dependent elastic tensor $C_{ijkl}(P)$ (see Supplementary Figure S1) determines the three acoustic velocities $V_P$, $V_{S1}$, and $V_{S2}$ for each propagation direction $\hat{n}$ through the Christoffel equation (Musgrave 1970; Wentzcovitch et al. 1998),

$$\left|C_{ijkl}n_jn_l-\rho V^2\delta_{ik}\right| = 0\,. \tag{3}$$

Here $C_{ijkl}$ is the elastic tensor, $n_j$ and $n_l$ are components of the unit propagation direction $\hat{n}$, $\rho$ is density, $V$ is acoustic velocity, and $\delta_{ik}$ is the Kronecker delta.

The directional velocity distributions at 300 K and various pressures are shown in Fig. 3. We uniformly sample propagation directions and summarize the resulting velocities using

$$f(V) = \frac{1}{N_{\mathrm{dir}}} \sum_{i=1}^{N_{\mathrm{dir}}} \delta(V - V_i) \, , \tag{4}$$

where $N_{\mathrm{dir}}$ is the number of sampled directions and $V_i$ is the velocity for direction $i$ and a specified acoustic branch. We refer to $f(V)$ as the "velocity density of states" (VelDOS). In practice, it is evaluated as a histogram. This distribution provides a compact summary of acoustic velocities across all propagation directions. For weakly anisotropic materials, the distribution is narrow, and one aggregate $V_P$ and $V_S$ may be sufficient. For lizardite, the distribution is broad and multi-peaked.

The VelDOS contains two prominent $V_P$ maxima and several $V_S$ features. The $V_S$ distribution includes a slow region associated with propagation along or near the principal crystallographic directions [001] and [100], together with two faster regions. We denote the off-axis secondary maximum by $V_S^*$; it arises from propagation directions intermediate between [001] and [100] and characterizes the intrinsic shear-wave anisotropy of lizardite. Similarly, the $V_P$ distribution contains a fast region and an off-axis secondary maximum, denoted by $V_P^*$. A smaller $V_P$ feature occurs close to $V_P^*$ and may not be readily distinguishable in the velocity distribution. The propagation directions associated with $V_P^*$ and $V_S^*$ differ and evolve differently with pressure (see Supplementary Figure S2).

The pressure dependence of the characteristic acoustic velocities is shown in Fig. 4. We track the evolution of characteristic features of the directional velocity distribution under pressure and compare them with VRH-based velocities and experimental measurements. For $V_S$, the fastest and slowest characteristic velocities show little pressure dependence, whereas $V_S^*$ exhibits a stronger pressure dependence between 0 and 2 GPa before becoming less pressure-sensitive at higher pressures. The pressure dependence of the characteristic $V_P$ velocities is generally stronger.

The comparison in Fig. 4 shows that the VRH-derived velocities differ substantially from the reported experimental velocities. The reported $V_P$ values track the slower off-axis $V_P^*$ region, whereas the reported $V_S$ values remain near the slow directional region associated with propagation close to [001]. Even the slower Reuss-bound velocities are substantially faster than the reported experimental velocities. This comparison indicates that ideal lizardite contains intrinsically slow propagation directions that are not captured by isotropic VRH averaging. The proximity of these slow directional velocities to the experimental values indicates that comparisons with experiment should retain directional information instead of reducing the calculation to a single isotropic aggregate velocity.

We also compare the PBE and LDA results (see Supplementary Figure S3). The precise peak positions depend on the functional, but multiple acoustic velocity peaks also occur in the PBE and LDA results. Their persistence across all three functionals identifies lizardite's strong elastic anisotropy as their structural origin. $r^2$SCAN improves the quantitative compression curve, whereas the multi-peak velocity distribution follows from the layered structure itself.

Sample texture and measurement geometry determine how an aggregate experiment samples the single-crystal directional velocity distribution. A randomly oriented aggregate samples a broad range of crystallographic directions, whereas a crystallographic or lattice-preferred orientation (CPO/LPO) may enhance some directional contributions relative to others. Ji et al. (2013) demonstrated how texture affects the measured anisotropy of antigorite-rich serpentinites. The antigorite results cannot be transferred quantitatively to lizardite, but they illustrate how intrinsic single-crystal anisotropy can affect rock-scale measurements. Conversely, the low-temperature lizardite/chrysotile-bearing samples studied by Watanabe et al. (2007) and Christensen (1966) showed relatively weak macroscopic anisotropy. Thus, texture is one of several factors that may affect the experimentally sampled velocities, but the central result here is independent of any particular texture model: ideal lizardite itself contains slow directional velocities that are obscured by isotropic VRH averaging.

## IMPLICATIONS

Lizardite's strong elastic anisotropy makes its seismic-velocity endmember dependent on orientational state. VRH averages for a randomly oriented aggregate, directional single-crystal velocities, and effective velocities of a textured aggregate are not interchangeable representations of $V_P$ and $V_S$. The VelDOS retains the breadth and modal structure of the directional velocity distribution that are collapsed into a single isotropic average.

This distinction directly affects velocity-based estimates of serpentinization and associated water content. Using the relatively high VRH velocities as the pure-lizardite endmember can attribute velocity reductions produced by slow crystallographic directions to additional serpentine, thereby overestimating serpentinization. Directional elasticity and texture should therefore be incorporated into rock-physics models used to interpret seismic velocities in serpentinized regions.

Measurements of $V_P$, $V_{S1}$, and $V_{S2}$ along multiple crystallographic directions would provide a direct experimental assessment of the calculated spectrum. Single-crystal measurements could locate the predicted principal-axis and off-axis velocity features, while measurements on aggregates with quantified lattice-preferred orientation would establish how these features contribute to the effective rock-scale response and connect the calculated elasticity of lizardite to seismic observations of serpentinized regions in cold subduction zones, where lizardite is the dominant low-temperature serpentine polymorph.

## ACKNOWLEDGMENTS

This work was supported by U.S. National Science Foundation (NSF) grant #2506448. R.M.W. acknowledges support from the Gordon & Betty Moore Foundation (doi.org/10.37807/GBMF12801). Calculations used allocation TG-DMR180081 through the Extreme Science and Engineering Discovery Environment (XSEDE) (Towns et al. 2014), which was supported by U.S. National Science Foundation (NSF) grant #1548562, and through the Advanced Cyberinfrastructure Coordination Ecosystem: Services & Support (ACCESS) program, which is supported by NSF grants #2138259, #2138286, #2138307, #2137603, and #2138296. Specifically, it used the *Bridges-2* system at the Pittsburgh Supercomputing Center (PSC), the *Anvil* system at Purdue University, and the Expanse system at San Diego Supercomputing Center (SDSC).

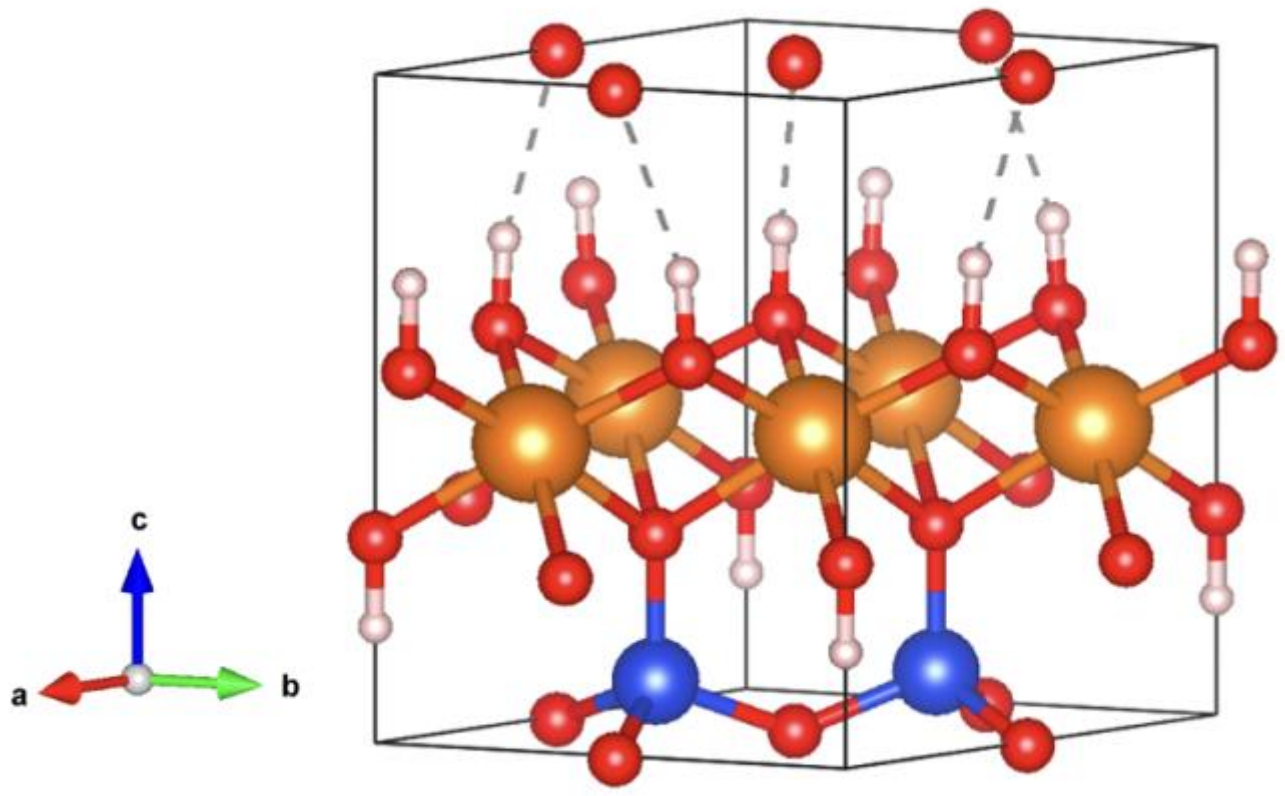

**Figure 1.** Lizardite unit cell. Mg, Si, O, and H atoms are shown in orange, blue, red, and pink, respectively.

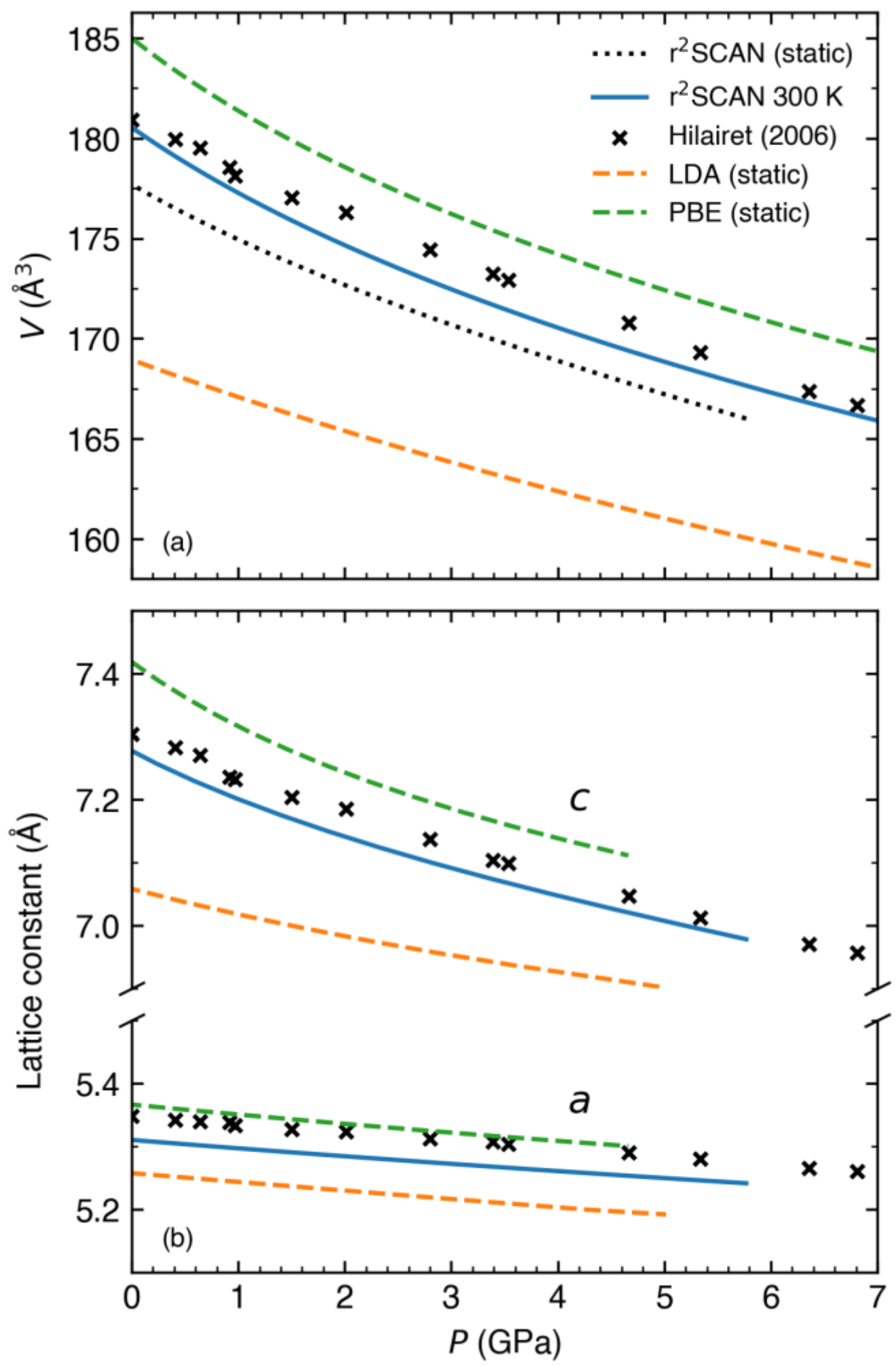

**Figure 2.** (a) Compression curves of lizardite. The dotted black line shows the $r^2$SCAN static result; the solid blue line shows the $r^2$SCAN 300 K result. Dashed orange and green lines show BM3 fits to static

LDA and PBE calculations, respectively. (b) Static lattice parameters *a* and *c* vs. pressure for r$^2$SCAN (blue), LDA (orange), and PBE (green). Black crosses in both panels show measurements from Hilairet et al. (2006).

**Table I.** Calculated volume, elastic constants, and aggregate velocities of lizardite at 0 GPa, together with selected reported experimental velocities at their stated pressures. Elastic constants are in GPa, volume is in Å$^3$, and velocities are in km/s. GULP denotes the General Utility Lattice Program; LT denotes low-temperature serpentinite containing lizardite and/or chrysotile. Source abbreviations combine the first author's initial with the two-digit publication year (e.g., D22 = Deng et al. 2022); full citations are given in the reference list.

| **Source** | **Method** | $V_0$ | $C_{11}$ | $C_{33}$ | $C_{12}$ | $C_{13}$ | $C_{14}$ | $C_{44}$ | $C_{66}$ | $V_P$ | $V_S$ |
|---|---|---|---|---|---|---|---|---|---|---|---|
| *Calculations* | | | | | | | | | | | |
| D22 | LDA, static | 172.54 | 227.33 | 124.74 | 82.28 | 27.27 | 0.38 | 16.24 | 72.53 | 7.31 | 3.89 |
| D22 | LDA, 300 K | 175.09 | 225.19 | 104.27 | 79.62 | 20.66 | 0.16 | 16.03 | 72.78 | 7.12 | 3.89 |
| D22 | GGA, static | 184.81 | 215.72 | 59.68 | 74.22 | 8.11 | 1.54 | 10.63 | 70.75 | 6.46 | 3.63 |
| G15 | LDA | 172.2 | 222.5 | 104.6 | 75.2 | 18.9 | 2.2 | 17.1 | 73.65 | 7.05 | 3.91 |
| MS09 | LDA | 170.76 | 235.61 | 118.16 | 85.96 | 25.05 | 2.69 | 20.92 | 74.83 | 7.41 | 4.08 |
| R07 | GGA | … | 245 | 23 | 50 | 31 | 0 | 11.6 | 97.5 | 6.20 | 3.66 |
| MS09 | GGA | 186.34 | … | … | … | … | … | … | … | … | … |
| T13 | GGA | 182.3 | 212.6 | 57.3 | 73.3 | 8.5 | 1.3 | 11.6 | 69.65 | 6.40 | 3.63 |
| A06 | GULP | 184.0 | 229.08 | 45.838 | 89.044 | 13.558 | 4.6025 | 12.765 | 70.02 | 6.50 | 3.67 |
| This study | r²SCAN, 300 K | 180.55 | 230.40 | 48.81 | 83.88 | 5.88 | 0.08 | 9.74 | 73.26 | 6.33 | 3.57 |
| *Experiments* | | | | | | | | | | | |
| H06 | XRD EoS, DAC | 180.92 | … | … | … | … | … | … | … | … | … |
| C66 | ultrasonic aggregate | … | … | … | … | … | … | … | … | 5.10 | 2.35 |
| C04/R07 | ultrasonic aggregate, extrap. 0 GPa | … | … | … | … | … | … | … | … | 4.70 | 2.26 |
| J13 | compiled LT-serpentinite endpoint, 0.6 GPa | … | … | … | … | … | … | … | … | 5.10 | 2.32 |
| W07 | ultrasonic LT serpentinite (95% Liz+Chr), 0.2 GPa | … | … | … | … | … | … | … | … | 5.03 | 2.62 |

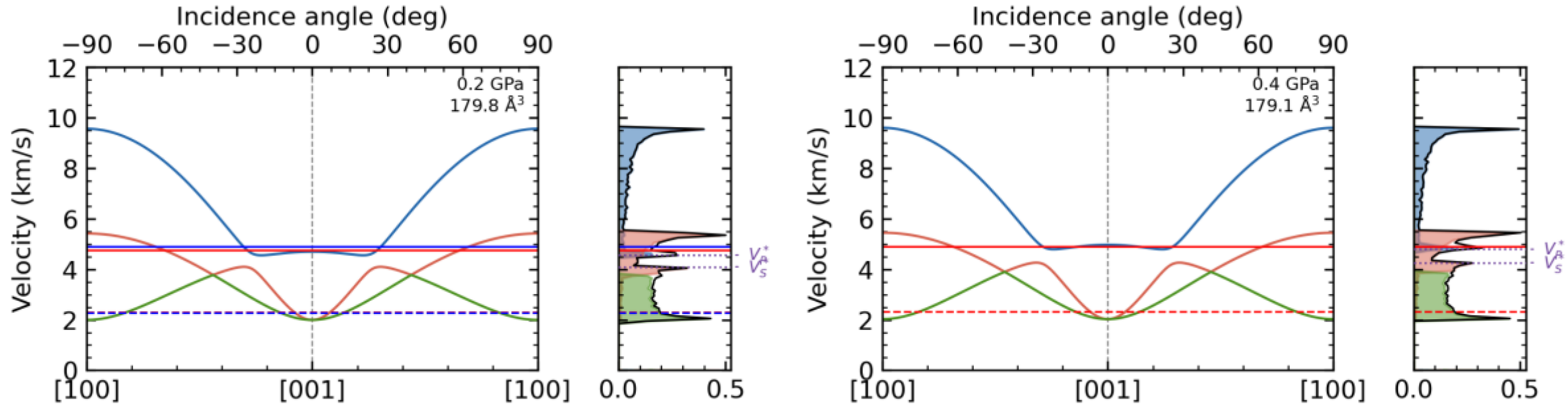

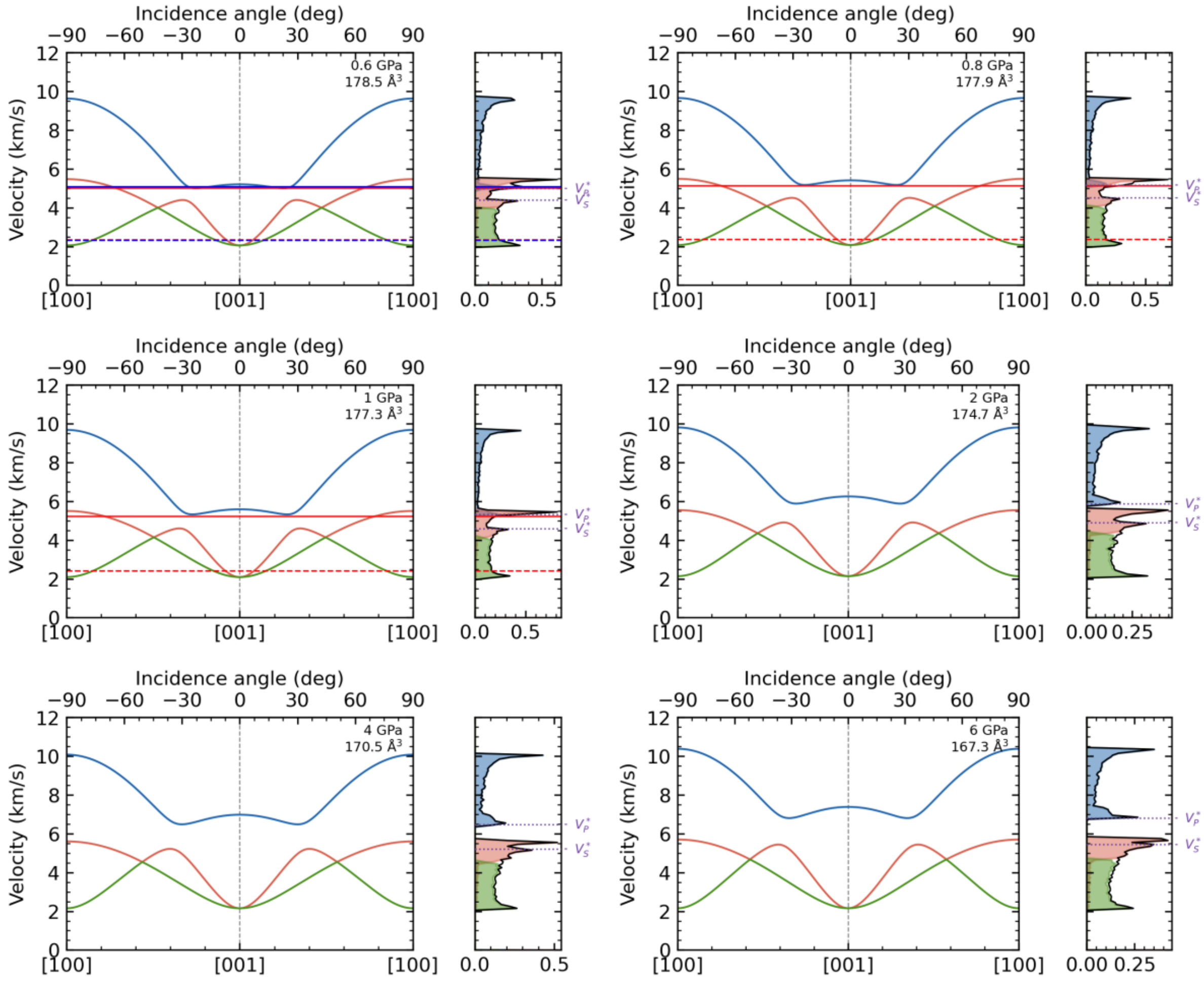

**Figure 3.** Directional distribution of lizardite's acoustic velocity at 300 K as a function of propagation direction along the [100]→[001]→[100] path. The blue, red, and brown areas show the distribution of $V_P$, $V_{S1}$, and $V_{S2}$, respectively (Fan et al. 2015). Horizontal lines mark reported experimental velocities, colored by source (red: Christensen (2004); blue: Ji et al. (2013)) and styled by wave type (solid: $V_P$; dashed: $V_S$). The right panels show the velocity density of states at each pressure. Dotted purple lines mark $V_P^*$ and $V_S^*$, characteristic off-axis local maxima arising from stationary regions of the directional velocity surfaces away from the principal crystallographic axes.

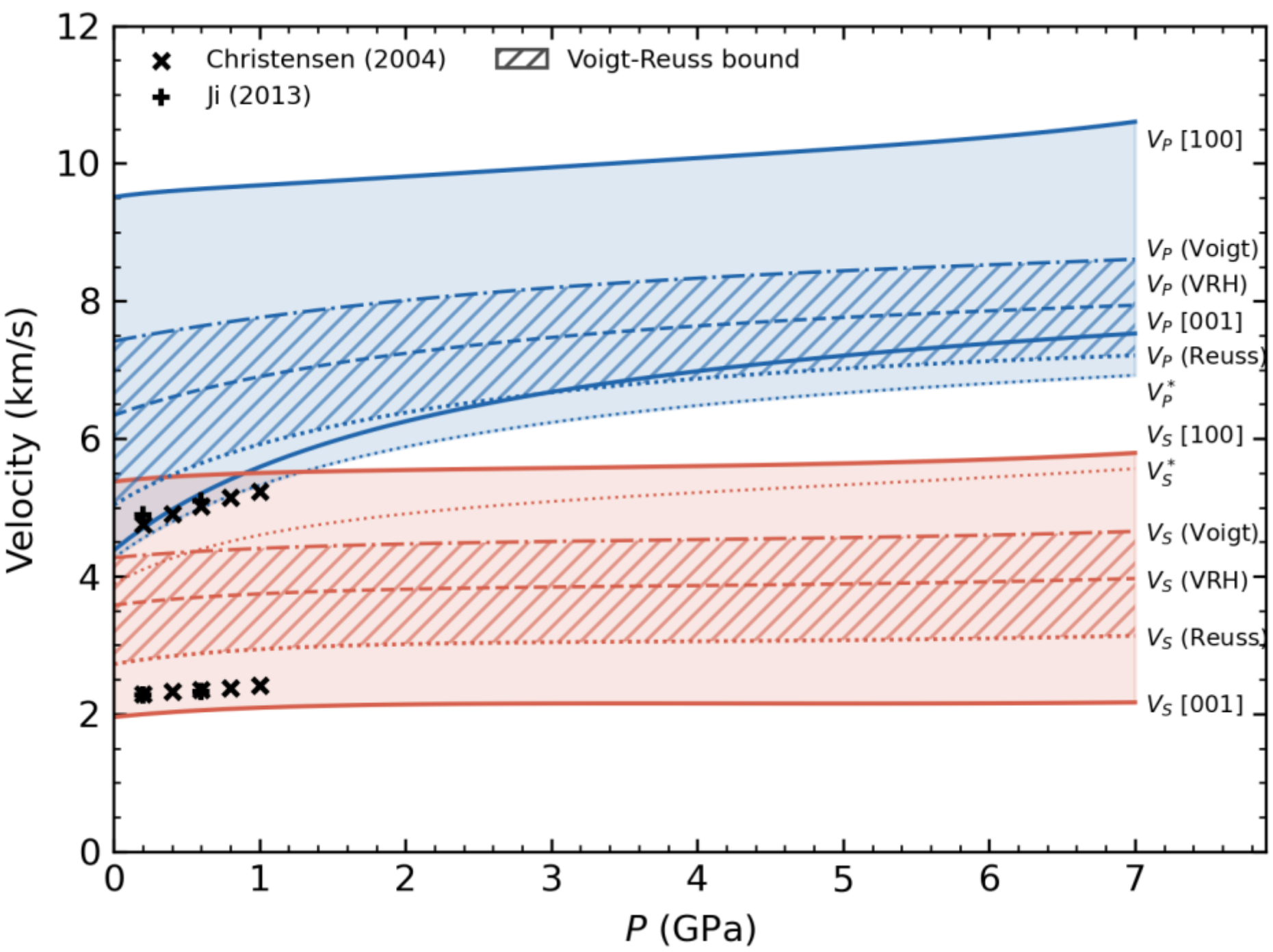

**Figure 4.** Acoustic velocities of lizardite at 300 K as a function of pressure. Solid curves show characteristic velocities along [100] and [001]. Dashed lines show $V_P$ and $V_S$ obtained from Voigt–Reuss–Hill (VRH) averaging. The semitransparent shading spans the full directional ranges for $V_P$ (blue) and $V_S$ (red), while the hatched regions indicate the Voigt–Reuss bounds. Symbols show experimental measurements: × from Christensen (2004) and + from Ji et al. (2013). $V_P^*$ and $V_S^*$ denote off-axis secondary maxima in the velocity density of states. The experimental $V_P$ values cluster near $V_P^*$, whereas the experimental $V_S$ values cluster near the slower region associated with propagation close to [001]; $V_S^*$ marks the faster off-axis maximum.